\documentclass[apjl]{emulateapj}

\usepackage{graphicx}
\slugcomment{accepted by ApJ Letters}

\shorttitle{Kinematics of HLX-1}
\shortauthors{Soria, Hau \& Pakull}

\begin{document}


\title{Kinematics of the intermediate mass black hole candidate HLX-1}


\author{Roberto Soria}
\affil{International Centre for Radio Astronomy Research, 
Curtin University, GPO Box U1987, Perth, WA 6845, Australia}
\email{roberto.soria@icrar.org}

\author{George K. T. Hau}
\affil{European Southern Observatory, Alonso de Cordova 3107, Santiago, Chile}
\email{ghau@eso.org}


\author{Manfred W. Pakull}
\affil{Observatoire Astronomique de Strasbourg, UMR 7550, Universit\'e 
de Strasbourg, 11, rue de l'Universit\'e, F-67000 Strasbourg, France}
\email{manfred.pakull@astro.unistra.fr}




\begin{abstract}
We studied the optical spectrum of HLX-1 
during its latest outburst, using the FORS2 spectrograph on the 
Very Large Telescope. We detect an H$\alpha$ emission line 
centered at $\lambda = 6718.9 \pm 0.9$ \AA\  and find that its  
projected radial velocity with respect to the nucleus of ESO\,243-49 
is $424 \pm 27$ km s$^{-1}$, 
while the maximum rotational velocity of the stars in that galaxy 
is $\approx 209$ km s$^{-1}$. This suggests that HLX-1 and its surrounding 
stars were not formed {\it in situ}, but came either from a disrupted 
dwarf galaxy or from a nuclear recoil.
We also find that the H$\alpha$ emission line is resolved with  
full width at half maximum $\approx 400$ km s$^{-1}$, suggesting 
a nebular rather than disk origin for the emission.
Its luminosity 
($L_{{\rm H}\alpha} \approx$ a few $10^{37}$ erg s$^{-1}$, equivalent width 
$\approx 70$ \AA) is also consistent with emission from a nebula 
photo-ionized by HLX-1. 
\end{abstract}


\keywords{Accretion, accretion disks --- black hole physics --- galaxies: individual (ESO\,243-49) --- galaxies: clusters: individual (Abell 2877)}



\section{Introduction}
The transient point-like X-ray source 2XMM\,J011028.1$-$460421 (HLX-1) 
has been interpreted as the first robust example of 
the long-sought class of intermediate-mass black holes (IMBHs) 
\citep{far09,wie10,dav11,ser11}.
HLX-1 is seen in the sky 
inside the D25 isophote of the S0/a galaxy ESO\,243-49 [heliocentric redshift 
$z = 0.0224$, cosmology-corrected \citep{ade13} 
luminosity distance $\approx 99$ Mpc, 
distance modulus $\approx 35.0$ mag], 
at a distance of $\approx 8\arcsec \approx 3.7$ kpc from the nucleus. 
Its peak X-ray luminosity $\approx 10^{42}$ erg s$^{-1}$ requires 
a BH mass $\ga$ a few $10^3 M_{\odot}$, to be consistent 
with the Eddington limit; this is too massive to be the outcome 
of a stellar collapse.  
This mass estimate is consistent with the values obtained 
from spectral modelling of the soft X-ray component, interpreted 
as thermal emission from an accretion disk \citep{far09,dav11,ser11}.
Its X-ray spectral variability and radio flares \citep{far09,god09,ser11,web12} 
are consistent with the canonical state transitions and jet properties 
of an accreting BH. 
The periodic nature of the X-ray outbursts (every $\approx 370$ d) 
suggests that the BH is accreting from a single donor star on a very 
eccentric orbit \citep{las11,sor13}.
Explaining the origin of HLX-1 may provide the key 
for finding further examples of IMBHs in the nearby universe, and 
for understanding their role in the history of galaxy evolution.

HLX-1 has a blue optical counterpart 
($B \sim V \sim 24$ mag near the outburst peak; \citealt{far12,sor12a,sor10})
with H$\alpha$ emission \citep{wie10}. We have studied the optical counterpart 
with the FORS2 spectrograph on the European Southern Observatory (ESO)'s 
Very Large Telescope (VLT) in 2012. In this paper, we present 
the main results of our observations and discuss their implications 
for the nature of HLX-1.

\section{Observations}

\subsection{Our 2012 VLT/FORS2 campaign}
We observed HLX-1 during its 2012 X-ray outburst \citep{god12},
with VLT/FORS2 (ESO Program 088.D-0974).
Spectra covering the R and I bands
were taken during the nights of 2012 August 26, 27 and September 10
(total exposure time $\approx 3.77$ hr) with the red-sensitive MIT CCDs.
We used the 300I grism with $1\arcsec$ slit and OG590 order-sorting filter.
The median seeing estimated from the telescope's active optics were
$\approx 0\farcs99, 0\farcs78 $ and $0\farcs70$ for the 3 nights
respectively. The observations were made
in blocks of $3\times754$-s exposures, with a spatial offset between
each exposure to facilitate cosmic-ray and bad pixel removal.
See Table 1 for
a summary of the instrumental configuration.

\begin{deluxetable}{lr}
\tablewidth{0pt}
\tablecaption{Instrument configuration for Program 088.D-0974(B)}
\tablehead{
\colhead{Parameter}           & \colhead{Value} }
\startdata
Epoch &    2012 Aug 27 (08:16--09:36 UT)\\[3pt]
    & 2012 Aug 28 (06:28--07:48 UT)\\[3pt]
    & 2012 Sep 11 (06:19--07:39 UT)\\[3pt]
Grism       & 300I \\[3pt]
Slit width & $1\farcs0$ \\[3pt]
CCD             & {\rm MIT}\\[3pt]
Read noise (e$^-$)   & 2.9\\[3pt]
Gain (e$^-$/ADU)      & 0.70\\[3pt]
Readout mode    & 100Kps/2ports/high\_gain\\[3pt]
Slit PA on sky   & $71^{\circ}.9$\\
\enddata
\label{tab:config}
\end{deluxetable}



Owing to its faintness, HLX-1 is not directly visible in the FORS2 acquisition 
images; therefore, we used an unresolved star-like object 
$9\farcs13$ from our target for alignment (Fig.~1),
and oriented the slit at a position angle of $71^{\circ}.9$ (from North 
to East). We calculated this value from our previously obtained, 
astrometrically-calibrated VLT/VIMOS images \citep{sor12a} and we verified it
using the public-archive {\it Hubble Space Telescope} ({\it HST}) 
Wide Field Camera 3 images. 
During our observations, the telescope was nodded along the slit 
to help us remove CCD artefacts.

\subsection{VLT/FORS2 2009 data}
We also re-analyzed the archival FORS2 data taken over 3 nights 
in 2009 December \citep{wie10}. The instrumental setup was similar 
to that used for our 2012 observations, except for the following details. 
The observations were a sequence of 600-s exposures (total exposure 
time $\approx 2$ hr). The alignment was done by placing the slit through 
two alignment stars, with a slit position angle of $-70^{\circ}.5$. 
Using a public-archive {\it HST} image, we note that with 
this position angle, HLX-1 was centered $\approx 0\farcs17$ 
off the middle of the slit, towards the blue side. For the seeing 
$\approx 0\farcs6$--$0\farcs7$ recorded on those nights, we estimate 
that this positional shift would cause an apparent blueshift 
$\sim 60$--$80$ km s$^{-1}$ of the H$\alpha$ emission component 
with respect to the stellar absorption component (which has no offset 
because the stellar emission fills the slit). Thus, combining the 2009 
and 2012 datasets increases the signal-to-noise ratio of the emission line, 
but also the error in the central position and line width. 
 

\section{Data Analysis}
We followed the standard steps for the reduction of long-slit spectral data 
(see e.g. \citealt{hau99,hau06,hau09}): the data were bias-subtracted 
and flat-fielded with a normalised internal flat. We obtained 
a wavelength solution using the sky emission lines in the individual spectra, 
ensuring optimal wavelength precision. The rms for the wavelength solutions 
is $0.8$ \AA.
We modelled and subtracted the background sky spectra by fitting the sky 
along the spatial direction column by column, excluding regions close 
to ESO\,243-49. We then spatially aligned and median-combined the data 
to reject pixels affected by cosmic rays. We created two separate 
2-D spectra from the coadded 2009 and 2012 observations, and another one 
with both datasets combined, to increase the signal-to-noise ratio 
of the emission line. We also inspected the H$\alpha$ line 
profiles from individual nights in 2012, but the signal-to-noise ratio 
is not high enough to determine whether the line parameters vary from 
night to night.

   \begin{figure}
   \centering
  \includegraphics[width=8.6cm]{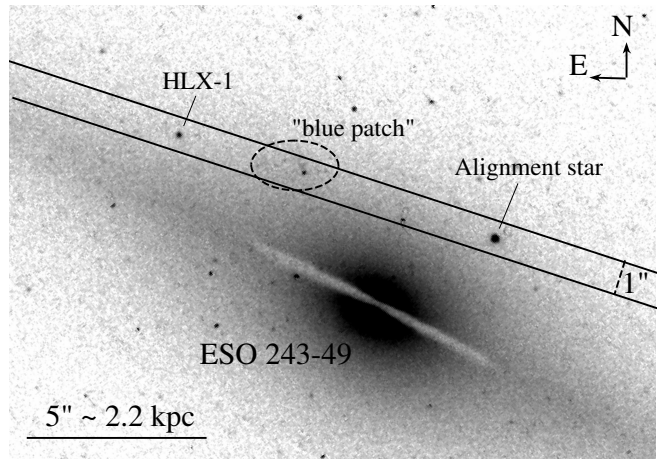}
   \caption{Slit geometry during our VLT/FORS2 observations, plotted on 
a greyscale {\it HST}/WFC3 image in the F390W band. The FORS2 spectrograph 
was rotated such that the $1\arcsec$-slit covered HLX-1, the alignment star, 
and what we called ``blue patch'', an extended background 
source particularly prominent in the far-UV.}
\vspace{0.3cm}
              \label{fig:acquisition}
    \end{figure}

The H$\alpha$ emission is directly visible in the combined 2-D spectra 
even before we subtract the stellar emission from ESO\,243-49 (Fig.~2, 
top panel). 
To obtain the galaxy-subtracted 2-D spectra of HLX-1, we subtracted a model 
of the stellar emission, by fitting a 2nd-order cubic spline
function with $2\sigma$ rejection 
along the spatial direction. The resulting spectrum is shown in Fig.~2 
(bottom panel); notice also the strong line emission from the candidate 
background galaxy (blue patch).
Finally, we obtained 1-D spectra of HLX-1 by summing over 
4 pixels ($1\arcsec$) 
in the spatial direction at each wavelength position along the slit (Fig. 3).

  \begin{figure}
   \centering
  \includegraphics[width=8.6cm]{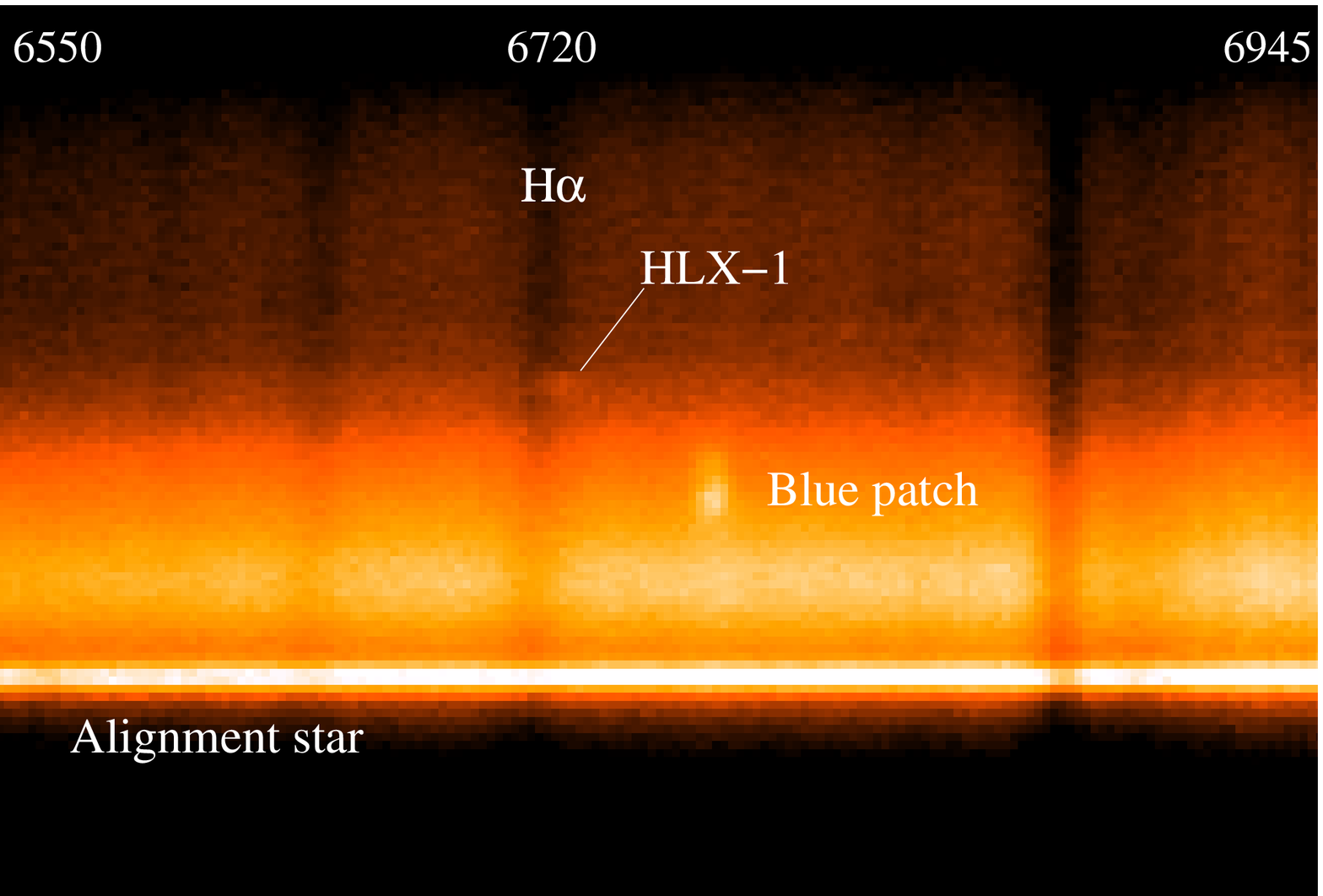}
  \includegraphics[width=8.6cm]{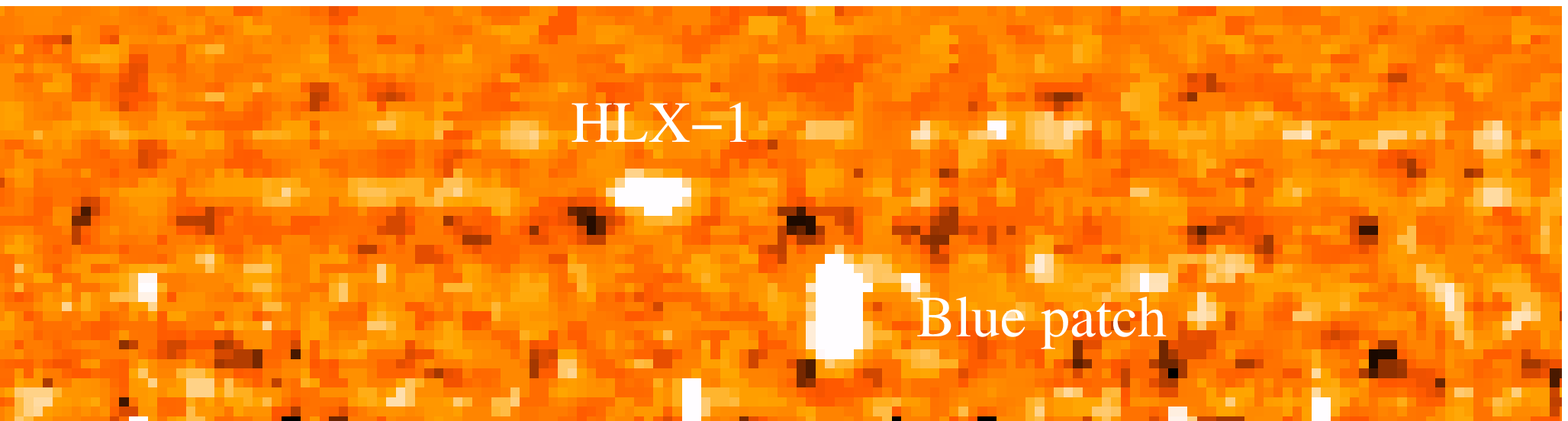}
   \caption{{\it Top panel}: 
Two-dimensional spectrum in the H$\alpha$ region (combined 2012 data), showing 
emission from HLX-1 and the candidate background galaxy (blue patch). 
The X axis is the wavelength direction, the Y axis the spatial direction. 
The wavelength range is $\lambda = 6550$--$6945$ \AA\ (as labelled at the top). 
An emission line from HLX-1 is apparent even before subtracting the stellar 
contribution from the halo of ESO\,243-29. Notice also that the HLX-1 emission 
feature is very close to but distinctly redshifted from the stellar 
H$\alpha$ absorption line. 
{\it Bottom panel}: zoomed-in view of the emission 
features, plotted on the same wavelength scale as the panel above, 
after we have subtracted a model of the stellar spectrum from ESO\,243-29.
The image has been smoothed over 3 pixels. H$\alpha$ from HLX-1 
is centered at $\lambda \approx 6718.9$ \AA\ (Section 4). 
The blue-patch line is unresolved (FWHM $< 7.5$\AA), centered 
at $\lambda \approx 6766.0$, corresponding to an apparent heliocentric 
recession velocity $\approx 9280$ km s$^{-1}$, and a true velocity 
$\approx 9200$ km s$^{-1}$ after we correct for the offset from 
the slit center towards the red side (Fig.~1).}
\vspace{0.3cm}
      \label{fig:red2d}
    \end{figure}

\section{Main results}

\subsection{Kinematics of HLX-1 and ESO\,243-49}
We confirm the presence of a strong emission line from HLX-1 (Fig.~3), 
as discovered by \citet{wie10}. We agree with \citet{wie10} 
that H$\alpha$ is the most plausible identification for this line, 
given the characteristic recession velocities of ESO\,243-49 and 
of the other galaxies in Abell 2877 \citep{mal92}. Henceforth, we shall assume 
this identification.
%
%
%
%
From the 2012 data, we measured a central wavelength 
$\lambda \approx 6718.9 \pm 0.5 \pm 0.8$ \AA; the errors are 
the statistical uncertainty and the systematic uncertainty 
in the wavelength solution (the latter is not relevant when we measure 
relative wavelength offsets). 
For the 2009 data, we measured  $\lambda \approx 6717.6 \pm 0.7 \pm 0.8$ \AA. 
This velocity difference ($\approx 60 \pm 40$ km s$^{-1}$) is almost 
certainly due to the offset position of HLX-1 in 
the slit in 2009, as expected. We take the 2012 measurement as the real 
line wavelength.
The corresponding heliocentric recession velocity 
of HLX-1 is $v \approx 7131 \pm 22 \pm 35$ km s$^{-1}$; 
that is a difference $\Delta v \approx 420$ km s$^{-1}$ 
with respect to the recession velocity of the nucleus of ESO\,243-49 
reported in the literature ($6714 \pm 34$ km s$^{-1}$: \citealt{cal97}).
The relative offset between the peaks of the H$\alpha$ line emission 
and (stellar) absorption components at the location 
of HLX-1 is $\approx 6$\AA.

To verify this interesting result, we determined the rotation curve 
of ESO\,243-49 along the slit, although our observational setup 
was not specifically designed for rotational velocity measurements. 
We started by summing spectra in $1\farcs25$ 
bins along the spatial direction. We then cross-correlated each of those 
galaxy spectra with a K2V template spectrum with solar metallicity 
(HD 149661: \citealt{val04}), using the {\it fxcor} task in the {\small {IRAF}} 
data analysis package. We restricted the cross-correlation to regions 
near the H$\alpha$ and near-IR Ca{\footnotesize{II}} triplets. 
The resulting rotation curve along the slit is shown in Fig.~4, 
where the centre is defined as where the slit crosses the projected galaxy axis.
The rotational velocity of ESO\,243-49 is $209 \pm 17$ 
km s$^{-1}$, measured as the difference 
between the recession velocities at $r=0$, and where the slit 
crosses the disk plane (at $r \approx 20\arcsec$).
With this method, we confirm a heliocentric systemic 
velocity $\approx 6707 \pm 16 \pm 35$ km s$^{-1}$ 
for ESO\,243-49, and we conclude 
that HLX-1 is offset from it by $424 \pm 27$ km s$^{-1}$; 
HLX-1 is receding at a speed $\approx 215$ km s$^{-1}$ faster 
than the peak stellar rotational velocity, and $\approx 270$ km s$^{-1}$ 
faster than the stellar population seen projected around it  
in the D25 of ESO\,243-49.

As a further check, we estimated the rotational velocity we should expect from 
a galaxy such as ESO\,243-49, based on the Tully-Fisher relation 
\citep{williams10,tully77}. 
The total absolute brightness (corrected for extinction and K-correction) 
of ESO\,243-49 is $M_B = -20.1 \pm 0.1$ mag, $M_{Ks} = -24.2 \pm 0.1$ mag 
(photometric measurements from the NED and Hyperleda databases). 
If we classify it as 
an Sa galaxy, the best-fitting Tully-Fisher relation gives a maximum 
rotational velocity $v_{\rm rot} = 210^{+30}_{-25}$ km s$^{-1}$ from 
the B-band brightness, or  $228^{+23}_{-22}$ km s$^{-1}$ 
from the $K_{s}$-band brightness. If we classify it as an S0 galaxy, 
the Tully-Fisher relation predicts a best-fitting 
$v_{\rm rot} = 244^{+35}_{-30}$ km s$^{-1}$, 
or $259^{+27}_{-25}$ km s$^{-1}$, for the two bands respectively.
Considering the large scatter around the best-fitting Tully-Fisher 
relations, the rotational velocity we measured along 
the slit is in agreement with these expected rotational velocities 
and can be taken as a good approximation of the maximum rotational velocity. 

The large velocity discrepancy proves that HLX-1 is kinematically 
decoupled from the stars at similar (projected) radial distances 
in ESO\,243-49. This suggests that HLX-1 and its surrounding star cluster 
were not formed {\it in situ}, but originated either from a captured 
satellite or as a gravitational recoil from the nuclear region.
We are unable to determine whether HLX-1 is bound or unbound 
to the galaxy---in the latter case it could be an intracluster IMBH.
Determining the radial profile of the escape velocity requires 
detailed modelling of the gravitational potential from the light profile 
and 2-D kinematics (which the current instrument setup is not designed for), 
and is beyond the scope of this paper; in any case, this would not 
provide a definitive answer 
because we only know the projected position and radial velocity.


   \begin{figure}
\vspace{-0.2cm}
\hspace{-0.3cm}
  \includegraphics[width=6.4cm, angle=-90]{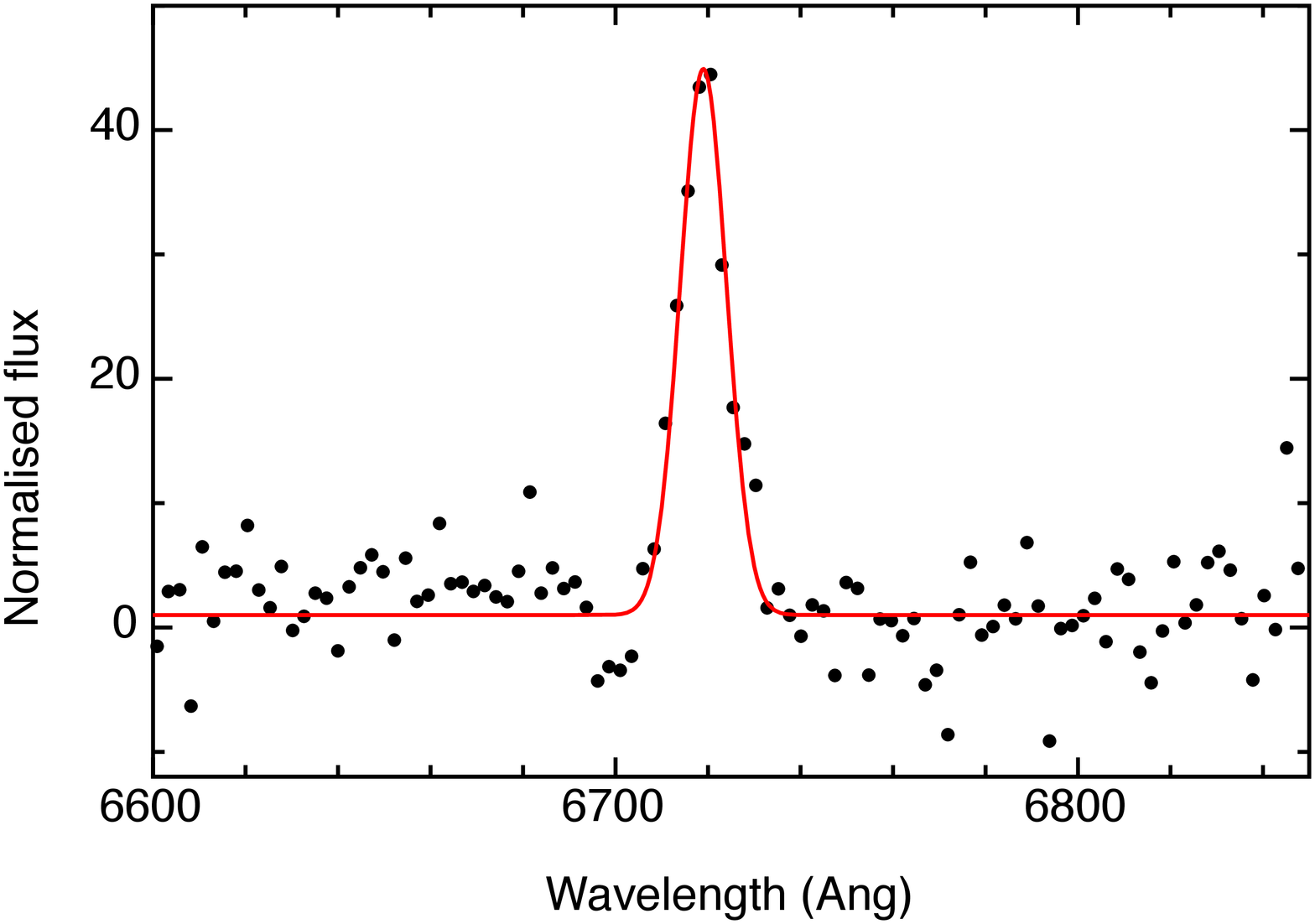}
   \caption{Normalized spectrum of the H$\alpha$ spectral region in 2012, 
after subtraction of the galactic halo component, with a Gaussian fit 
to the emission line. The spectrum has been rebinned to 
a resolution of $2.44$\AA/pixel.}
\vspace{0.5cm}
              \label{fig:hlx1_1d}
    \end{figure}

  \begin{figure}
\vspace{-4.1cm}
\hspace{-5.9cm}
  \includegraphics[width=14.3cm, angle=-90]{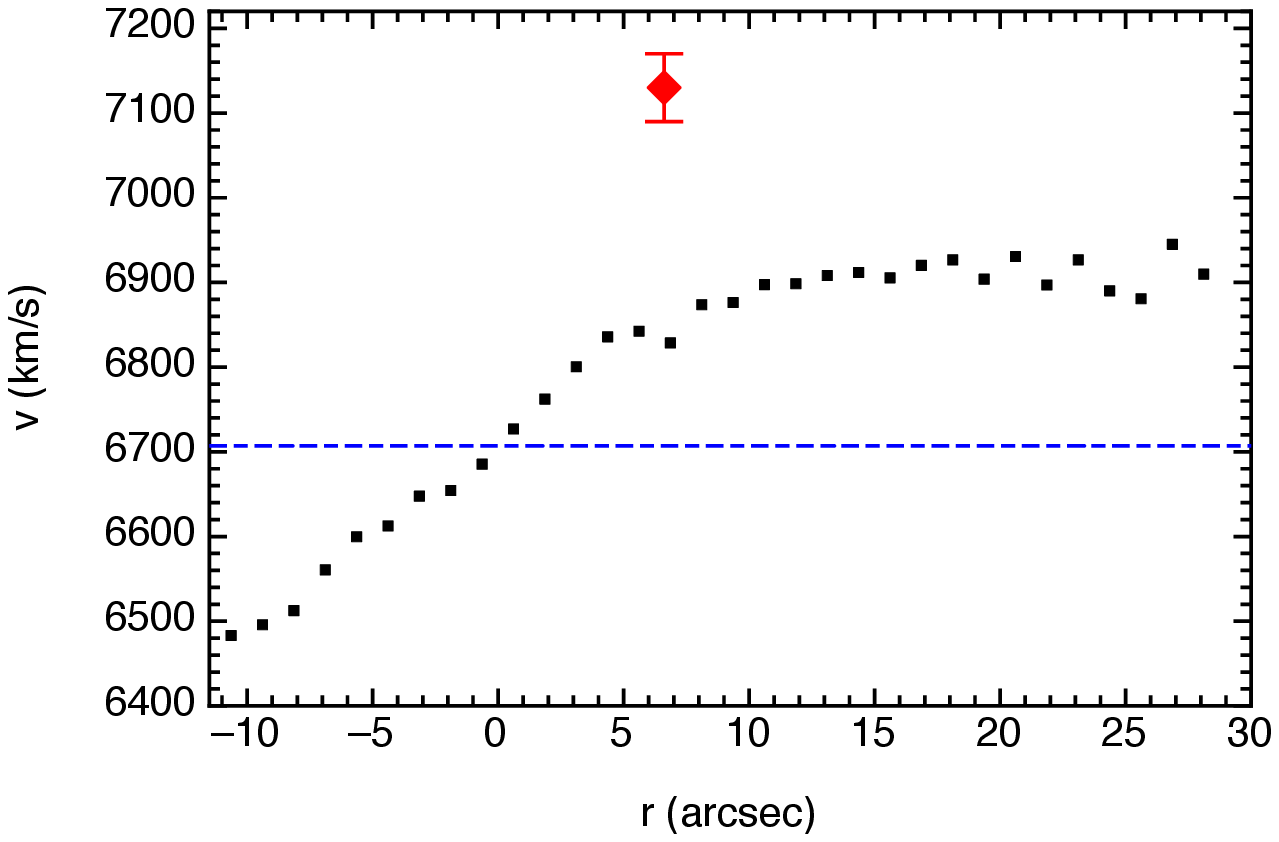}
\vspace{-4.1cm}
   \caption{Rotation curve of ESO\,243-49 along the slit, 
with the 2012 recession velocity of HLX-1 indicated by the red dot. 
The systemic velocity of ESO\,243-49 is indicated by the blue line. 
On the X axis, $r$ increases from West to East.}
              \label{fig:}
\vspace{0.5cm}
    \end{figure}

\subsection{Width and flux of the H$\alpha$ line}
From the combined, background-subtracted 2012 spectrum, we measured a Gaussian 
full-width-half-maximum (FWHM) $=12.5 \pm 1.0$ \AA\ for 
the H$\alpha$ emission line.
The instrumental linewidth inferred 
from the median seeing\footnote{From the sky lines, we determined that 
the fully illuminated slit gives a FWHM resolution 
of $\approx 9.9$ \AA; for a point-like source in the centre of the slit, 
this value is then linearly scaled by the seeing, 
when it is $< 1\arcsec$.} is $\approx 7.8$ \AA; deconvolving it from 
the observed FWHM, we obtain that the line is resolved 
with an intrinsic FWHM $\approx (9.8 \pm 1.0)$ \AA\ $\approx 440$ km s$^{-1}$.
However, when we consider only the average of the two nights with best seeing 
(2012 August 28 and September 11), we obtain a FWHM $= 10.3 \pm 1.2$ \AA\
for a seeing-limited instrumental linewidth $\approx 7.4$ \AA;
this corresponds to an intrinsic FMHM 
$= 7.2 \pm 1.2$ \AA\ $\approx 320$ km s$^{-1}$.

We estimate an integrated line flux of $(0.62 \pm 0.05) e^{-}$ s$^{-1}$
in the combined 2012 spectrum.
A faint continuum trace in the 2-D spectra is visible by eye 
in a few exposures taken with excellent seeing ($\la 0\farcs7$).
We mean-combined three 754-s exposures with the best signal-to-noise ratio
(two from 2012 September 11 and one from 2012 August 28). We extracted
the continuum from two spectral regions relatively free of interstellar
lines, and excluding the H$\alpha$ line itself: from $6420$--$6680$ \AA\ and
from $6740$--$7065$ \AA; we averaged the measured values from the two bands,
and obtained a continuum level of 
$(22 \pm 5) e^{-}/{\rm pixel} \approx (6.8 \pm 1.5) e^{-}/$\AA.
This corresponds to an equivalent width (EW) of H$\alpha$ $ = (70 \pm 15)$ \AA.

To convert the instrumental fluxes into physical units, we used
the FORS2 Exposure Time Calculator (ETC).  For a blackbody spectrum 
at $T \approx 15,000$--$20,000$ K, the continuum corresponds
to $R = 24.5 \pm 0.5$ mag (Vega). The line flux is 
$f_{{\rm H}\alpha} = (2.0 \pm 0.3) \times 10^{-17}$ erg cm$^{-2}$ s$^{-1}$, 
corresponding to a luminosity $\approx 2.3 \times 10^{37}$ erg s$^{-1}$
(consistent with \citealt{wie10}).

Instead of using the somewhat uncertain (because of the rapidly 
changing background along the slit) optical flux calibrations,
we could {\it assume} that the R-band continuum was the same as measured
by {\it HST}/WFC3 on 2010 September 23, that is $R \approx 23.7$ (Vegamag)
$\approx 23.9$ (ABmag) (from \citealt{far12,map13}, interpolated between
F555W and F814W). In that case, for our measured EW, the line flux is 
$f_{{\rm H}\alpha} = (4.5 \pm 1.0) \times 10^{-17}$ erg cm$^{-2}$ s$^{-1}$,
corresponding to a luminosity $\approx 5.2 \times 10^{37}$ erg s$^{-1}$.
However, there is still no proof that the optical/UV continuum 
flux is constant from
epoch to epoch, as it is still debated whether it has a significant component
from the irradiated accretion disk.

Galactic stellar-mass BHs often display H$\alpha$ emission from the outer 
annuli of their accretion disks (e.g., \citealt{fra02}), 
with typical EW $\la 10$\AA\ in the high-soft 
state and EW $\sim 10$--$100$\AA\ during the decline towards quiescence 
\citep{fender09}.  
The exact value of the FWHM depends on the radial profile of the emissivity 
function \citep{sma81}; 
however, for typical disk sizes and temperatures, 
if $V_d$ is the rotational speed of the outermost annulus 
and $i$ is the viewing angle, 
a useful first-order approximation is FWHM $\approx 2 V_d \sin i$. 
In a Keplerian approximation, $V_d = \left(GM/R\right)^{1/2}$. 
For a BH mass $\sim 10^4 M_{\odot}$ and an outer disk radius $\la 10^{13}$ cm 
\citep{dav11,far12,sor12a,sor13}, we expect $V_d \ga 3500$ km s$^{-1}$ 
and FWHM $\ga 7000 (\sin i)$ km s$^{-1}$. Therefore, the narrow 
FWHM of the line observed from HLX-1 either requires an extremely face-on disk 
($i \la 3^\circ$) or, more likely, implies that the line does not 
come from a Keplerian disk. It might come, for example, from a larger  
X-ray photoionized nebula (XIN) around the BH, perhaps similar to those 
seen around some ULXs
(for example, RZ 2109 in NGC\,4472: \citealt{pea12}; 
Holmberg II X-1: \citealt{pak02,kaa04,leh05}; 
NGC\,5408 X-1: \citealt{kaa09}) 
or stellar-mass BHs (LMC X-1: \citealt{pak86}).
Observations of those XINs, and models of line luminosities based 
on the photo-ionization code {\small {CLOUDY}} \citep{fer13},
suggest that the H$\alpha$ luminosity of an XIN is $\approx 0.3\%$  
of the ($0.3$--$10$) keV luminosity of the X-ray source. 
The exact value depends, among other parameters, on the detailed spectral 
energy distribution of the X-ray source and the nebular density, and should 
be considered to be indicative only within a factor of three. With this 
reservation in mind, we find that the H$\alpha$ emission from HLX-1 
could be excited by an X-ray luminosity of $\sim 1 \times 10^{40}$ erg/s, 
which is an order of magnitude smaller than the mean luminosity 
observed over 2009--2012.
Possibly, the interstellar density is very small at the position 
of the X-ray source, or the emitting material has a low filling factor, 
or, alternatively, the mean luminosity of the source has indeed been  
lower in the past. Note that the H$\alpha$ luminosity reflects 
the long-term-average X-ray luminosity over the recombination timescale 
$\tau = 1/\left[n_{e}\,\alpha(H^0,T)\right] \sim \left(10^5/n_e\right)$ years, 
where $\alpha$ is the hydrogen recombination coefficient at temperature $T$, 
and $n_e$ the electron density \citep{ost89}.



\section{Conclusions}
With our optical spectroscopic study, 
we have shown that HLX-1 has a high velocity offset $\Delta v_1 \approx 420$ 
km s$^{-1}$ with respect to the galactic nucleus, 
and $\Delta v_2 \approx 270$ km s$^{-1}$
with respect to the stellar rotational velocity 
at its projected location in the halo of ESO\,243-49.
We have also shown that the H$\alpha$ emission line is resolved 
with a FWHM $\approx 400$ km s$^{-1}$, a luminosity 
$\approx$ a few $10^{37}$ erg s$^{-1}$ and an EW $\approx 70$ \AA. 

The kinematics of HLX-1 proves that it was not formed {\it in situ}: it is 
either the stripped remnant of a satellite dwarf 
(consistent with the simulations of \citealt{map12,map13}), 
or a recoiling BH ejected from the nucleus of ESO\,243-49, 
dragging along a small, compact cluster of stars and gas inside 
its sphere of influence \citep{hof06,mer09}. 
In the former case, it would have a very elongated, almost parabolic  
orbit (if bound at all to ESO 243-49).
In the latter case, it may be unbound and destined to become soon 
(within $\sim$ few $10^7$ yrs) a free-floating intracluster IMBH.
However, the recoiling BH scenario would be ruled out 
if most of the optical/UV emission came from a massive star cluster 
\citep{far12} rather than an accretion disk.

Based on the large relative velocity of HLX-1, 
we speculate that there must exist other active IMBHs similar to HLX-1 
not located or projected inside the D25 of a galaxy at the moment, 
either because they are on very eccentric orbits with 
large semimajor axes, or because they have been ejected from a galaxy. 
Some of them may already have been observed in X-ray surveys but 
were perhaps misidentified and dismissed as background quasars, 
in the absence of any deep optical studies. 
We propose that a clue to identify those sources is an X-ray 
to optical flux ratio $> 100$ (typical of IMBHs or stellar-mass BHs) 
associated with a soft X-ray spectrum and blue optical colors. 

The small FWHM of H$\alpha$ suggest that it does not 
come from the irradiated surface of a standard accretion disk. 
It is more likely to come from ionized gas at larger distances from the BH.
This could be warm gas remaining in a young star cluster (age $\la 5$ Myr), 
or a nebula or outflow photo-ionized by the IMBH, 
similar to some ULX nebulae found in nearby galaxies.
Deep long-slit observations in the blue region of the optical spectrum would 
provide a key test for the presence of other emission  
lines such as $[$\ion{O}{3}$]$ $\lambda 5007$ and \ion{He}{2} $\lambda4686$; 
in particular, the latter would provide strong indication of a ULX bubble.


\acknowledgments
RS acknowledges a Curtin University Senior Research Fellowship, 
and the hospitality of Marie-Claude Moery and of the Strasbourg Observatory.
GKTH thanks ESO for the Director General's Discretionary Fund grant 12/18/C 
for this project, and the ESO staff who supported these observations. 
He also thanks the hospitality of Curtin University of Technology during 
his visit. We thank the referee for an insightful review, 
and Giovanni Carraro, Sean Farrell, Jeanette Gladstone, 
Alister Graham, James Miller-Jones, Michela Mapelli, 
Ivo Saviane, Luca Zampieri for comments and suggestions.

\clearpage

\clearpage

\end{document}